\documentstyle[aps,prd,preprint]{revtex}

\def\beq {\begin{equation}}
\def\eeq {\end{equation}}

\begin{document}
\draft


\title{Quantum dynamics of non-relativistic particles and isometric embeddings}

\author{Alberto Saa}

\address{Departamento de Matem\'atica Aplicada,\\ 
IMECC -- UNICAMP,\\
C.P. 6065, 13081-970 Campinas, SP, Brazil}


\maketitle

\begin{abstract}
It is considered, in the framework of constrained systems, 
the quantum dynamics of non-relativistic particles moving
on a $d$-dimensional Riemannian manifold $\cal M$ isometrically
embedded in $R^{d+n}$.
This generalizes recent investigations where $\cal M$ has been 
assumed to be a hypersurface of $R^{d+1}$. We show, contrary to
recent claims, that constrained systems theory does not contribute
to the elimination of the ambiguities present in the canonical and
path integral formulations of the problem. These discrepancies with
recent works are discussed.
\end{abstract}
\begin{flushleft}
PACS: 0350, 0365, 0420
\end{flushleft}

\newpage

\section{Introduction}

Some recent works
have been devoted to the study of non-relativistic
particles in curved spaces in the framework of constrained 
systems\cite{OFK,HIM,OCK,O,SIM,FGK}.
The idea is that the theory of constrained systems\cite{GT,HT}
 might shed some light in the
long standing problem of the quantization of non-relativistic particles in
curved spaces\cite{DW}. Such a 
problem consists in a non-relativistic point particle
of mass $M$ 
moving on a $d$-dimensional Riemannian manifold $\cal M$ and 
is described 
by the Lagrangian $L = \frac{M}{2}g_{\alpha\beta}\dot{q}^\alpha\dot{q}^\beta$.
The tangent vectors $\dot{q}^\alpha = \frac{d}{dt} q^{\alpha}$ are the
velocities of the particle and $g_{\alpha\beta}$ is the metric tensor of 
$\cal M$. The Lagrangian $L$ is non-singular, in contrast to the relativistic
case, and the Hamiltonian equations of motion can be immediately obtained
\begin{eqnarray}
\label{ham}
\dot{q}^\alpha &=& \{q^\alpha, H \}, \nonumber \\
\dot{p}_\alpha &=& \{p_\alpha, H \}, \nonumber \\
H &=& \frac{1}{2M} g_{\alpha\beta}(q)p^\alpha p^\beta,
\end{eqnarray}
where $p_\alpha$ are the momenta canonically conjugated to $q^\alpha$.
It is clear that the determination of $\hat{H}$,
the quantum counterpart of $H$ 
necessary to define the Schr\"odinger equation, is plagued with severe
ordering ambiguities, which represent the greatest difficulty of the 
problem\cite{DW}. From the correspondence principle, the unitarity of
the quantum evolution, and dimensional analysis,
 one concludes that $\hat{H}$
must have the form
\beq
\hat{H} = - \frac{\hbar^2}{2M} \frac{1}{\sqrt{g}}\partial_\alpha 
\sqrt{g}g^{\alpha\beta}\partial_\beta + \kappa\frac{\hbar^2}{M}R, 
\eeq
where $R$ is the scalar of curvature of $\cal M$.
There is no consensus in the literature about the dimensionless constant
$\kappa$. Different approaches have led to different values for $\kappa$
(Reference \cite{FGK}, for instance,
 contains a recent survey on the subject.). The path integral
formulation does not solve these ambiguities. It is possible to read off
the operator $\hat{H}$ directly from the phase space path integral of the
quantum evolution operator $K$ of the problem\cite{B}, but then
the value of
$\kappa$ will
depend on the parameterization used to evaluate the 
$K$ path integral.

In the works \cite{OFK,HIM,SIM,FGK} it is considered the
case where $\cal M$ is a hypersurface of $R^{d+1}$. The
motion of the particle in $R^{d+1}$ 
is enforced to take place on $\cal M$ by the
introduction of a Lagrange multiplier. $\hat{H}$ is then obtained in
the canonical and path integral formulations by following the
standard steps of the theory of constrained systems\cite{GT,HT}.
Although it is well known that
the Hamiltonian operators obtained from the canonical and
path integral formulations may not coincide in general\cite{M},  
it is claimed that $\kappa$ could be set unambiguously for each formulation
in the constrained systems framework.

The case where $\cal M$ is a hypersurface of $R^{d+1}$ is an
 especial and convenient one because of
the existence of an unique (up to a sign) normal vector in all points of
$\cal M$. However,
we know that even 
locally the hypersurfaces of $R^{d+1}$ correspond only to
a small portion of the $d$-dimensional manifolds for $d>2$. 
In order to see it, 
consider the embedding of a $d$-dimensional Riemannian manifold
 $\cal M$ in $R^{d+1}$. $\cal M$
can be described parametrically by the equations
\beq
\label{emb}
x^i = x^i(q^\alpha), \ \ \ \{i=1,...,d+1; \alpha=1,...,d \},
\eeq
such that the rank of the matrix 
$B^i_\alpha = \frac{\partial x^i}{\partial q^\alpha}$ is maximal. From 
(\ref{emb}) we have that, for $dx^i$ tangent to $\cal M$, 
$ds^2=\eta_{ij}dx^i dx^j=\eta_{ij}B^i_\alpha B^j_\beta dq^\alpha dq^\beta$, 
where 
$\eta_{ij}=\eta^{ij}={\rm diag}(1,...,1)$. The embedding is called
isometric if
the metric tensor $g_{\alpha\beta}$ of $\cal M$ is given by
\beq
\label{eqa}
g_{\alpha\beta} = \eta_{ij}B^i_\alpha B^j_\beta.
\eeq
 A metric tensor in a 
$d$-dimensional manifold has $\frac{d(d+1)}{2}$ independent components, and
we see
the right hand side of (\ref{eqa}) has only $d+1$ independent components.
By this simple argument, we have that for $d>2$  equation (\ref{eqa}) defines
only a restricted class of metrics. The solution is to raise the number of 
independent components of the right hand side  by raising
the dimension of the Euclidean space where $\cal M$ is embedded. Indeed,
it is possible to embed isometrically
 any orientable Riemannian manifold in an Euclidean space of sufficiently
high dimension\cite{K}. In the references \cite{OCK,O}, 
the canonical quantization
for the case of $\cal M$ isometrically embedded in a higher dimensional
Euclidean space was considered.

It is quite surprising that for the case of $\cal M$ isometrically embedded,
one gets for $\hat{H}$ an extra contribution proportional to the
square of the extrinsic mean curvature of $\cal M$\cite{OFK}. 
Such a geometrical quantity is not intrinsic to
$\cal M$ and it also appears
when the particle is enforced to move on $\cal M$ due to external
potentials (See \cite{DMO} for a recent approach and for references).

The purpose of the present
work is to show that constrained systems theory cannot
contribute to the solution of the ambiguities of the problem, 
and in particular, it does not give a definite answer
about the value of $\kappa$. We 
perform the Hamiltonian analysis of a particle moving in a
Riemannian manifold $\cal M$ isometrically
embedded in a high-dimensional Euclidean space.
Such a system has a set of second--class constraints, and we 
show that there is
a {\em canonical} transformations casting the dynamical equations 
for the physical variables in
the form (\ref{ham}). 
The canonical quantization of theories with second--class constraints
in different canonical variables leads to equivalent physical 
theories\cite{GT}, and thus
all ordering ambiguities inherent to (\ref{ham}) are also present
in the constrained system framework, and the difficulties to solve
them are essentially the same ones of the usual analysis.
We will see that although the physical
Hamiltonian will be apparently free of ordering ambiguities,
they will be hidden in the constraints of the problem.
By using the same canonical transformation, we will show that also the
constrained path integral formulation has exactly the same
ambiguity problems of the usual analysis. Our results are
compatible with the idea that Hamiltonian and Lagrangian descriptions
of singular systems are fully equivalent\cite{GT}.

The next section is devoted to the canonical formulation of the
problem. In Sect. \ref{III} we discuss the path integral approach, and
the last section is left to some concluding remarks.
 
\section{Canonical formulation}

Consider  a particle of mass $M$ 
moving in $R^{d+n}$, $n>0$. We can enforce the motion of
the particle to take place on a $d$-dimensional submanifold $\cal M$ of
$R^{d+n}$ by
imposing $n$ constraints $f^A(x) = 0$. Greek indices
run over ($1,...,d$), lower case roman ones over ($1,...,d+n$), and
upper case ones over ($1,...,n$). The Lagrangian of the particle
moving on $\cal M$ is
\beq
\label{lagr}
L = \frac{M}{2} \eta_{ij} \dot{x}^i\dot{x}^j + \lambda_Af^A(x),
\eeq
where $\lambda_A$ are $n$ Lagrange multipliers. The functions $f^A(x)$
must be non-degenerated in the sense that the $n$ covariant vectors
$\frac{\partial f^A}{\partial x^i}$ are linearly independents for all
points of $\cal M$. Such a condition assures that the 
tangent space of $\cal M$
has $(d+n)-n=d$ dimensions for all points of $\cal M$.
The Lagrangian $L$ is singular  and when going to the Hamiltonian formalism
we get constraints. In this case the first-stage constraints are
\beq
\label{c1}
\Phi_1^A = P^A = 0,
\eeq
where $P^A$ are the momenta canonically conjugated to the
variables $\lambda_A$. We can now construct the hamiltonian 
$H^{(1)}$
\beq
H^{(1)} = \frac{1}{2M} \eta^{ij} \pi_i \pi_j - \lambda_Af^A(x) + 
\xi_A\Phi^A_1,
\eeq
where $(x^i;\pi_j)$ are canonical variables and $\xi_A$ another multipliers.
By following Dirac procedure, we check for the
existence of $(m+1)$-stage constraints by verifying the conservation in time
of the $m$-stage ones. We have the following new constraints
\begin{eqnarray}
\label{c2}
\Phi_2^A &=&  \dot{\Phi}_1^A = f^A(x) = 0, \nonumber \\
\Phi_3^A &=& M\dot{\Phi}_2^A = \pi^i \partial_i f^A = 0, \\
\Phi_4^A &=& M\dot{\Phi}_3^A = \pi^i\pi^j \partial_i\partial_j f^A
+ M\lambda_B S^{BA}=0, \nonumber 
\end{eqnarray}
where $S^{BA} = \eta^{ij} \partial_i f^B\partial_j f^A$. The
condition $\dot{\Phi}_4 = 0$ determines $\xi_A$ and no more constraints
arise. One can check by using elementary properties of determinants that
$S^{BA}$ is non-singular if $f^A$ are non-degenerated. The constraints
(\ref{c1}) and (\ref{c2}) form a set of second--class ones, 
in contrast to the relativistic case, where it is well known that due to the
reparameterization invariance we also have first--class constraints.
Furthermore,
$\Phi_1$ and $\Phi_4$ are constraints
of special form\cite{GT}, and they can be used
to eliminate the variables $\lambda_A$ and $P^A$. There remain
$2(d+n)$ variables and $2n$ constraints, what shows that in fact the system
has $2d$ degrees of freedom. The classical equations
\begin{eqnarray}
\label{hec}
\dot{x}^i &=& \{x^i, \bar{H} \} , \nonumber \\ 
\dot{\pi}_i &=& \{\pi_i, \bar{H} \} , \nonumber \\ 
\Phi_2^A &=& f^A(x) = 0, \nonumber \\
\Phi_3^A &=& \pi^i \partial_i f^A(x) = 0, 
\end{eqnarray}
with $\bar{H} = \frac{1}{2M}\eta_{ij}\pi^i\pi^j - \lambda^A f_A$,
are dynamically equivalent to (\ref{ham}). Note that due to such a
 choice for
$\bar{H}$, the symplectic structure of the phase space $(x^i;\pi_j)$ 
is just the usual Poisson brackets. The Hamiltonian
$\left.\bar{H}\right|_{\Phi=0}$ 
appears to be free of ordering ambiguities, and thus a
good starting point for canonical quantization. However, it indeed has
the ambiguities entangled to those ones of the constraints. 
In order to see it, we will perform a canonical
transformation such that the constraints will be free of ordering
ambiguities.

Our canonical transformation will be essentially a coordinate transformation.
In order to construct it, first
 note that due to the assumption that the vectors
$\partial_i f^A$ are linearly independent, the neighbourhood of $\cal M$ 
is spanned by the coordinates $(q^\alpha, f^A)$, where $q^\alpha$ are
coordinates in the submanifold ${\cal M}_W$ of
$R^{d+n}$ defined by $f^A$ constants,
{\em i.e.}
\beq
f^A(x(q)) = W^A
\eeq
identically for constants $W^A$. The vectors 
$B^i_\alpha = \frac{\partial x^i}{\partial q^\alpha}$ and 
$\partial_i f^A$ span respectively the tangent and normal spaces of
${\cal M}_W$. The canonically conjugated variables
$(x^i;\pi_j)$ and $(q^\alpha, f^A ; p_\beta, \theta_B )$ are related by
the following canonical transformation
\begin{eqnarray}
\label{eq1}
x^i &=& x^i(q,f) , \nonumber \\
\pi_i &=& \frac{\partial q^\alpha}{\partial x^i} p_\alpha +
          \frac{\partial f^A}{\partial x^i} \theta_A, \nonumber \\
\pi^i &=& B^i_\alpha p^\alpha + \frac{\partial x^i}{\partial f^A}\theta^A.
\end{eqnarray}
In the new canonical variables
$(q^\alpha, f^A ; p_\beta, \theta_B )$, the constraints $\Phi_2$ and
$\Phi_3$ are given by the following ambiguity-free expressions
\begin{eqnarray}
\Phi_2^A &=& f^A = 0 , \nonumber \\
\Phi_3^A &=& \theta^A = 0.
\end{eqnarray}
On the other hand,
$(q^\alpha, p_\beta)$ obey the usual equations (\ref{ham}).
This finally shows that the canonical quantization  based in
(\ref{hec}) and in (\ref{ham}) are plagued by the same ordering ambiguities.
In the equations (\ref{hec}), the ordering ambiguities are clearly 
in the constraint $\Phi_3$, which involves $\pi_i$ and a function of
$x^i$. The constraint $\Phi_3$  has many inequivalent and equally
acceptable operatorial representations. In the references \cite{OFK,OCK,FGK},
for instance, it is used a symmetric expression for the 
operatorial representation of $\Phi_3$, and it is obtained $\kappa=0$
and a contribution proportional to extrinsic geometrical quantities.
In the other hand, in the reference \cite{HIM} another choice for
the operators lead to $\kappa=0$ and no extrinsic contributions.
We could yet
choose another representation compatible with the requirement of
hermiticity of the theory. The constraint $\Phi_3$ is used explicitly in order
to determine $\hat{H}$, and different representations would lead
to distinct $\hat{H}$.

\section{Path-integral formulation}
\label{III} 
The derivation of the path integral for system with second--class
constraints is straightforward\cite{HT}.
Note first that 
due to that the symplectic structure of the phase space $(x^i,\pi_j)$
is given by the usual Poisson brackets, the Liouville measure of 
$(x^i;\pi_j)$ will be simply 
$\prod_{i=1}^{d+n} dx^i\, d\pi_i$. Thus,
the quantum
evolution operator $K$ of the system governed by (\ref{lagr}) will be given
by
\beq
K= \int D\lambda\, DP\, Dx\, D\pi \delta\left(\Phi_1\right)
\delta\left(\Phi_2\right)\delta\left( \Phi_3\right)
\delta\left(\Phi_4\right) \exp \left\{ 
\pi_i \dot{x}^i - \frac{1}{2M} \pi_i\pi^i + \lambda_Af^A - \xi_A\Phi^A_1
\right\}.
\eeq
The integration over the variables $\lambda$ and $P$ is straightforward
and allows us to write $\lambda$ as a function of the variables
$(x^i;\pi_j)$.
After doing it, we have 
\beq
\label{p1}
K=\int Dx\, D\pi\, \delta\left(f^A(x)\right)
\delta\left( \pi^i\partial_i f^A(x)\right)
\exp\left\{
\pi_i\dot{x}^i - \frac{1}{2M} \pi_i\pi^i + \lambda_A f^A
\right\}.
\eeq
The integration over the momenta and constraints using the midpoint
parameterization allows us to get the Weyl ordered Hamiltonian
operator directly from (\ref{p1})\cite{B}. This was done for the
case of $\cal M$ embedded in $R^{d+1}$ in \cite{SIM,FGK}, and it was
obtained a non vanishing $\kappa$ and extra extrinsic terms.
However, 
we can check that in the new variables $(q^\alpha, f^A ; p_\beta, \theta_B )$
(\ref{p1}) reads
\beq
\label{p2}
K = \int Dq\, Dp\, \exp \left\{ 
p_\alpha \dot{q}^\alpha - \frac{1}{2M} p_\alpha p ^\alpha
\right\}.
\eeq
The expression (\ref{p2}) is the same one we would obtain starting from
the system described by (\ref{ham}), and thus it has exactly the same
ambiguities problems. The results of \cite{SIM,FGK} are a
consequence of the choice of the phase space variables and thus 
one cannot claim that the Weyl ordered $\hat{H}$ can be determined
unambiguously in the path integral formulation of the problem.

\section{Final remarks}

To summarize, we have shown that constrained systems theory does not
contribute to an unambiguous description of the quantum dynamics of
non-relativistic particles in curved spaces. This is not a great
surprise since we know that constrained systems theory does not solve
the analogous ambiguities of the relativistic case. For such a case, 
we have, besides of the ambiguities described here, the subtle issue
of time-depending constraints\cite{GG,ET}. However, it is possible
to avoid the latter by assuming a static spacetime, and recently we
have shown that the ambiguities questions remain unsolved in this case\cite{S}.
In the works \cite{OFK,HIM,OCK,O,SIM,FGK}, 
the values for $\kappa$ are obtained
by using implicit assumptions about the ordering of $\Phi_3$ in
(\ref{hec}) and the parameterization used to evaluate (\ref{p1}). Such
assumptions are sometimes privileged ones due to the phase space
coordinates used,
but nevertheless they are arbitrary and not unique. It is clear that
canonical and path-integral quantizations do not commute with
canonical transformations in the classical phase space.

We finish noticing that the use of $\dot{f}^A(x)=0$ instead of
$f^A(x)=0$ as the constraints 
to enforce the particle to move on $\cal M$ does not
lead to any improvement in the ordering ambiguities. Such a
possibility was first tried for the case of $R^{d+1}$ in
\cite{HIM}. In our case, starting with
\beq
L = \frac{M}{2} \eta_{ij} \dot{x}^i\dot{x}^j + \lambda_A \dot{f}^A
\eeq
instead of (\ref{lagr}), would lead to the following Hamiltonian
equations
\begin{eqnarray}
\label{hec1}
\dot{x}^i &=& \left\{ x^i, H' \right\} , \nonumber \\
\dot{\pi}_i &=& \left\{ \pi_i, H' \right\} , \nonumber \\
\Phi_1^A &=& P^A = 0 , \nonumber \\
\Phi^A_2 &=& \eta^{ij}\partial_i f^A \left( 
\lambda_B \partial_j f^B - \pi_i
\right) = 0,
\end{eqnarray}
where $H' = \frac{\eta^{ij}}{2M}\left(\pi_i - \lambda_A \partial_if^A \right)
\left(\pi_j - \lambda_A \partial_jf^A \right) + \xi_A P^A$, and
$\xi_A$ is determined from the condition $\dot{\Phi}_2^A=0$.
The quantization of the system governed by (\ref{hec1}) will
be also plagued with severe ordering ambiguities. Although we can use
$\Phi^A_1$ and $\Phi^A_2$ to eliminate the variables $P^A$ and
$\lambda_A$, the resulting Hamiltonian will have several ordering
dependent terms. The path integral formulation is equivalent to
the case discussed in the Sect. \ref{III}.

\acknowledgments

This work was supported by FAPESP.


\begin{references}

\bibitem{OFK}N. Ogawa, K. Fujii, and A. Kobushkin, Progr. Theor. Phys.
{\bf 83}, 894 (1990).

\bibitem{HIM}T. Homma, T. Inamoto, and T. Miyazaki, Phys. Rev. {\bf D42},
2049 (1990).

\bibitem{OCK}N. Ogawa, N. Chepilko, and A. Kobushkin, Progr. Theor. Phys.
{\bf 85}, 1189 (1991).

\bibitem{O}N. Ogawa, Progr. Theor. Phys. {\bf 87}, 513 (1992).

\bibitem{SIM}A. Shimizu, T. Inamoto, and T. Miyazaki, Nuovo Cim. {\bf B107},
973 (1992).

\bibitem{FGK}A. Foerster, H.O. Girotti, and P.S. Kuhn, Phys. Lett.
{\bf 195A}, 301, (1994).

\bibitem{GT} D.M. Gitman and I.V. Tyutin, {\em Quantization of fields with
constraints}, Springer-Verlag, 1990.

\bibitem{HT} M. Henneaux and C. Teitelboim, {\em Quantization of gauge
systems}, Princeton University Press, 1992.

\bibitem{DW}B.S. DeWitt, Phys. Rev. {\bf 85}, 653 (1952).

\bibitem{B} F.A. Berezin, Theor. Math. Phys. {\bf 6}, 1994 (1971).

\bibitem{M}M.S. Marinov, Phys. Rep. {\bf 60C}, 1 (1980).

\bibitem{K} S. Kobayashi and K. Nomizu, {\em Foundations of
differential geometry}, Vol. 2, Note 18, John Wiley, 1969.

\bibitem{DMO} D. Destri, P. Maraner, and E. Onofri, Nuovo Cim. {\bf 107A},
237 (1994).

\bibitem{GG} S.P. Gavrilov and D.M. Gitman, Class. Quant. Grav. 
{\bf 10},  57 (1993).

\bibitem{ET} J.M. Evans and P.A. Tuckey, {\em Geometry and
dynamics with time dependent constraints}, in {\em Geometry of
constrained dynamical systems}, Ed. J.M. Charap,
Cambridge Univ. Press, 1995. 

\bibitem{S} A. Saa, Class. Quant. Grav. {\bf 13}, 553 (1996).
 
\end{references}
\end{document}